\documentclass[fleqn,usenatbib]{mnras}
\usepackage{gensymb}

\usepackage{newtxtext,newtxmath}

\usepackage[T1]{fontenc}

\DeclareRobustCommand{\VAN}[3]{#2}
\let\VANthebibliography\thebibliography
\def\thebibliography{\DeclareRobustCommand{\VAN}[3]{##3}\VANthebibliography}

\usepackage{graphicx}	
\usepackage{amsmath}	
\usepackage{comment}

\title[Chorus]{Chorus: Optimizing Synchrotron Transfer Coefficients with Weighted Sums}

\author[D. van Duren, M. Mo{\'s}cibrodzka]{
D. van Duren,$^{1}$\thanks{E-mail:david.vanduren@ru.nl}
M. Mo{\'s}cibrodzka,$^{1}$\thanks{E-mail:m.moscibrodzka@astro.ru.nl}\\
$^{1}$Department of Astrophysics/IMAPP, Radboud University,P.O. Box
  9010, 6500 GL Nijmegen, The Netherlands
}

\date{Accepted 2025 June 4. Received 2025 May 30; in original form 2025 March 28}

\pubyear{2025}

\begin{document}
\label{firstpage}
\pagerange{\pageref{firstpage}--\pageref{lastpage}}
\maketitle

\begin{abstract}
Accurate synchrotron transfer coefficients are essential for modeling radiation processes in astrophysics. However, their current calculation methods face significant challenges. Analytical approximations of the synchrotron emissivity, absorptivity, and rotativity are limited to a few simple electron distribution functions that inadequately capture the complexity of cosmic plasmas. Numerical integrations of the transfer coefficients, on the other hand, are accurate but computationally prohibitive for large-scale simulations. In this paper, we present a new numerical method, Chorus, which evaluates the transfer coefficients by expressing any electron distribution function as a weighted sum of functions with known analytical formulas. Specifically, the Maxwell-Jüttner distribution function is employed as the basic component in the weighted sum. The Chorus leverages the additivity of transfer coefficients, drawing inspiration from an analogous approach that uses stochastic averaging to approximate the $\kappa$ distribution function.
The key findings demonstrate median errors below $5\%$ for emissivity and absorptivity, with run times reduced from hours to milliseconds compared to first-principles numerical integrations. Validation against a single $\kappa$ distribution, as well as its extension to more complicated distributions, confirms the robustness and versatility of the method. However, limitations are found, including increased errors at higher energies due to numerical precision constraints and challenges with rotativity calculations arising from fit function inaccuracies. Addressing these issues could further enhance the method's reliability. Our method has the potential to provide a powerful tool for radiative transfer simulations, where synchrotron emission is the main radiative process. 
\end{abstract}




\section{Introduction}

Synchrotron radiation is relevant in many fields of astrophysics, providing an exceptional mean for understanding the properties of matter in our universe. 
Common sources of synchrotron radiation include neutron stars and pulsars, as well as supernova remnants, which are considered to play a crucial role in shaping the chemical composition of the interstellar medium \citep{wilson1992abundances}. Furthermore, active galactic nuclei, X-ray binaries, and $\gamma-$ray bursts all containing accretion disks and relativistic jets, are sources of synchrotron radiation, emitting substantial energy across a wide range of wavelengths \citep{petrov2025radio, malzac2016jet}. All these sources share a common feature: the presence of magnetic fields and electrons moving at relativistic speeds.

In order to gain a deeper understanding of astrophysical objects of interest, it is essential to implement synchrotron radiation into models. One commonly used technique is ray-tracing simulations, which provide researchers with more freedom in choosing objects and allowing a simpler approach for finding breaking points in modern theories, without being subject to constraints created by observational limitations. Synchrotron radiation is emitted by relativistic electrons and interacts with plasma through absorption, Faraday rotation (rotation of polarization), and conversion  (conversion of linear polarization to circular and vice versa). In the radiative transfer equation, these interactions are represented by the transfer coefficients called the emissivity $j_{\nu}$, absorptivity $\alpha_{\nu}$ and the rotativity $\rho_{\nu}$. Although scattering processes, which formally are part of radiative transfer equation, can also be an important radiative process in some astrophysical environments, it is not included in the present work. The central dependence of all transfer coefficients is the electron distribution function $f(\gamma)$, which describes how electrons are distributed across an energy spectrum, represented as the Lorentz factor $\gamma$.

Matter in astrophysical objects is constantly evolving due to dynamic processes such as accretion, explosions, and magnetic interactions. Each object experiences unique environmental conditions, such as variations in temperature, density, and magnetic field strength. These factors influence how energy is distributed among the electrons, leading to different electron distribution functions. For example, electrons in a supernova remnant may follow a non-thermal distribution with significant high-energy content due to shock acceleration \citep{margalit2024peak}, whereas electrons in cooler environments, such as molecular clouds, are more likely to follow thermal distributions that peak at lower energies \citep{lowe2022study}. However, it has been shown that evaluating the transfer coefficients of the electron distribution function is both time-consuming and computationally intensive \citep{schwinger1949classical}. 

Currently, there are two options for calculating transfer coefficients for an arbitrary electron distribution function. The first option involves using a small set of electron distribution functions with analytical expressions, which are fit functions to the numerical results \citep{pandya2016polarized}. Analytical expressions are fast to compute but may be inaccurate. The second option involves using a numerical integration method (e.g., Symphony code,  \citealt{marszewski2021updated}). In theory, this method can be applied to any arbitrary electron distribution function. However, in its current state, it is too slow for practical use in more complicated simulations, where transfer coefficients often need to be evaluated hundreds of thousands to millions of times. Therefore, a more efficient method is required that is capable of accurately evaluating transfer coefficients for any arbitrary electron distribution function within a fraction of a second.

In this paper, a new method is proposed for evaluating synchrotron transfer coefficients for arbitrary electron distribution functions, designed to be computationally efficient and suitable for use in ray-tracing simulations.
The method is based on the observation made by \citet{mao2017impact} that synchrotron transfer coefficients exhibit an additive property. For example, synchrotron emissivity can be expressed as $j_{\nu} (f_1 + f_2) = j_{\nu} (f_1) + j_{\nu} (f_2)$ which implies that any arbitrary electron distribution function can be represented as a weighted sum of distribution functions with known, approximate analytical expressions. \citet{moscibrodzka2024stochastic} have recently shown that the relativistic $\kappa$ distribution function can be constructed from the Maxwell-Jüttner distribution functions and derived the analytical expression for the weights in the summation process. The current paper will generalize their approach in two key steps. First, an expression is determined for the weights, as they are the only unknown variables in the approximation. Our approach involves minimizing an error function with respect to the weights, where the error function incorporates the target distribution function and the approximation formula using the weights. Minimization can also be achieved by taking the derivative of the function with respect to the weights and setting it to zero. This yields an expression for the weights that need to be solved.
Second, before solving newly found expression for the weights, it is necessary to impose a non-negativity constraint on the weights. Negative weights could result an unphysical negative distribution function, which in turn could result in unphysical negative transfer coefficients. A widely used method in data science for solving such constrained problems is quadratic programming. Quadratic programming involves minimizing a function while incorporating constraints, either directly within the minimization function or by applying them to a numerical solver. For this project, the interior-point solver Clarabel was used, which is developed by \citet{goulart2024clarabel}, proved to be the most effective solution for this problem.

In this paper, the solver will first be applied to approximate a $\kappa$ distribution function using $N$ Maxwell-Jüttner distribution functions (often referred to as a relativistic thermal distribution function), each assigned a specific weight. Since both the $\kappa$ and Maxwell-Jüttner distribution functions have analytical expressions, they provide an ideal test case for our method.
The resulting approximation will then be used to reevaluate the transfer coefficients, and consequently will be compared to the aforementioned numerical integration using Symphony \citep{pandya2016polarized} in terms of both accuracy and efficiency. Furthermore, since the method is designed to handle arbitrary distribution functions, it will also be tested on a sum of three distinct $\kappa$ functions forming a more complicated distribution function.

\section{Preliminaries}

\subsection{Synchrotron radiation}

Synchrotron light is emitted from charged particles, such as electrons, moving at relativistic speeds through magnetic field. The particles feel the Lorentz force from the magnetic field and are forced to move in a circular path. As this particle is constantly changing its direction of movement, it experiences acceleration, which results in emission of electromagnetic waves. This motion causes the particle to lose energy, but also produce beams of radiation that spans a broad range of frequencies, depending on the energy of the particle and the strength of the magnetic field.

Relativistic beaming causes this radiation to be emitted in a narrow cone along the particle's direction of motion, making the emission highly directional and highly polarized, with the apparent intensity depending on the observer's angle relative to the particle's path.

\subsection{Radiative transfer equation and transfer coefficients}

The radiative transfer equation is the basis of many ray-tracing simulations. This equation describes the change of specific intensity $I_{\nu}$ along a spacial parameter $s$, where we neglect scattering. The radiative transfer equation for unpolarized light is:
\begin{equation}
    \frac{dI_{\nu}}{ds} = j_{\nu} - \alpha_{\nu}I_{\nu}
    \label{simple rad. trans}
\end{equation}
where $I_{\nu}$ is the specific intensity at a frequency $\nu$, $j_{\nu}$ is the emission coefficient, and $\alpha_{\nu}$ is the absorption coefficient. These two terms, $\alpha_{\nu}$ and $j_{\nu}$, incorporate the most significant interactions of radiation with the medium, although they do not capture all possible effects such as Faraday rotation and conversion, which will be discussed next. A key insight from Equation~\ref{simple rad. trans} is that, at each point along the radiation's path, the transfer coefficients must be reevaluated based on the local conditions of the matter. 

Synchrotron radiation is polarized, and incorporating polarization provides valuable insights into the magnetic fields of the medium of interest. The radiative transfer equation for polarized light is:
\begin{equation}
    \frac{d}{ds}\begin{bmatrix} I_{\nu} \\ Q_{\nu} \\ U_{\nu} \\ V_{\nu}\end{bmatrix} = \begin{bmatrix} j_{\nu , I} \\ j_{\nu , Q} \\ j_{\nu , U} \\ j_{\nu , V}\end{bmatrix} - \begin{bmatrix} \alpha_{\nu, I} & \alpha_{\nu, Q} &\alpha_{\nu, U}& \alpha_{\nu, V} \\ \alpha_{\nu, Q} & \alpha_{\nu, I} & \rho_{\nu, V} & -\rho_{\nu, U} \\ \alpha_{\nu, U} & -\rho_{\nu, V} & \alpha_{\nu, I} & \rho_{\nu, Q} \\ \alpha_{\nu, V} & \rho_{\nu, U} & -\rho_{\nu, Q} & \alpha_{\nu, I}\end{bmatrix} \begin{bmatrix} I_{\nu} \\ Q_{\nu} \\ U_{\nu} \\ V_{\nu}\end{bmatrix}
    \label{stokes rad. trans}
\end{equation}
where specific intensity $I_{\nu}$ and $j_{\nu}$ are now extended into a vectors expressed in terms of the Stokes parameters $IQUV$. Additionally, the expanded equation incorporates the Mueller matrix, which replaces the $\alpha_{\nu}$ with a matrix that includes absorptivities $\alpha_{\nu}$ and the rotativity $\rho_{\nu}$  (Faraday rotation and conversion coefficient) across different Stokes parameters. 

\begin{figure}
    \centering
    \includegraphics[width=1\linewidth]{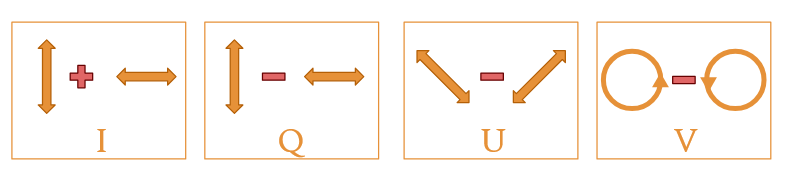}
    \caption{Stokes parameters illustrating the specific intensity associated with different types of light polarization.}
    \label{fig:stokes}
\end{figure}

The Stokes parameters $Q_{\nu}$, $U_{\nu}$ and $V_{\nu}$ describe the polarization state of light, with $Q_{\nu}$ and $U_{\nu}$ representing linear polarization at orientations typically defined relative to a reference axis in the plane perpendicular to the wave propagation. $V_{\nu}$ corresponds to circular polarization. A simplified visual representation is shown in Figure~\ref{fig:stokes} to aid intuition; for a complete and rigorous definition, see \cite{sault1996understanding}

Synchrotron rotativity $\rho_{\nu}$ describes both Faraday rotation and Faraday conversion. Faraday rotation refers to the rotation of the plane of linear polarized light in the presence of a magnetic field, resulting in an exchange between Stokes $Q_{\nu}$ and $U_{\nu}$. Faraday conversion, on the other hand, describes how linear polarized light is converting to circular polarized light and vice versa.

In Equation~\ref{stokes rad. trans}, one might notice that eleven different transfer coefficients is needed for every step in parameter $s$. As shown by \cite{pandya2016polarized} and \cite{leung2011numerical}, the equation for polarized synchrotron emissivity coefficients for an isotropic distribution is
\begin{equation}
    j_{\nu, S} = \begin{bmatrix} I_{\nu} \\ Q_{\nu} \\ U_{\nu} \\ V_{\nu}\end{bmatrix} = \frac{2\pi e^2 \nu^2 n_e}{c} \int d^3p f (p)
    \sum^{\infty}_{n=1} \delta (y_n) K_S 
    \label{emissivity gen.}
\end{equation}
where $p$ is the momentum and $f (p) \equiv  (1/ n_e) dn_e/d^3p$ the distribution function per particle, equivalently the distribution function can also be written as $f (\gamma) \equiv  (1/n_e)dn_e/d\gamma$, where $\gamma$ is the Lorentz factor and assuming an isotropic distribution function\footnote{While methods for computing transfer coefficients using anisotropic distribution functions have been developed, these evaluations are generally much slower (e.g. \citet{verscharen2018alps}).}. Also, $S = I,Q,U,V$, $\delta$ is the Dirac delta function, $K_S$ and $y_n$ are functions described in e.g. \cite{leung2011numerical}. Both $\alpha_{\nu, S}$ and $\rho_{\nu, S}$ follow similar equations to Equation~\ref{emissivity gen.}, but the most important aspect is that these coefficient are all linearly dependent on the electron distribution function $f(\gamma)$.

In practice, Equation~\ref{emissivity gen.} is computationally intensive and time-consuming to integrate. To address this, \cite{leung2011numerical} and \cite{pandya2016polarized} integrate this equation numerically and provide fit functions for emissivities and absorptivities, with solutions derived for thermal, $\kappa$ and power-law distribution functions, respectively. In the case of rotativity, both \cite{dexter2016public} and \cite{marszewski2021updated} derived fit functions for the thermal and kappa distribution functions. In this work, these fit functions will be used to demonstrate the results from Chorus.

Symphony is a C code developed by \citet{pandya2016polarized} that generates the numerical transfer coefficients used to derive the previously mentioned fit functions. While these numerical results are highly accurate, they are time-consuming to compute, making them impractical for direct use in simulations. Nonetheless, they provide a valuable benchmark for the method developed in this work.

\subsection{Electron distribution function}
The electron distribution function, $f (\gamma)$, describes the distribution of electrons in a medium as a function of the Lorentz factor, $\gamma$. Since $\gamma$ is directly related to energy, it can also be interpreted as representing the distribution of electrons across an energy spectrum.

\begin{figure}
    \centering
    \includegraphics[width=1\linewidth]{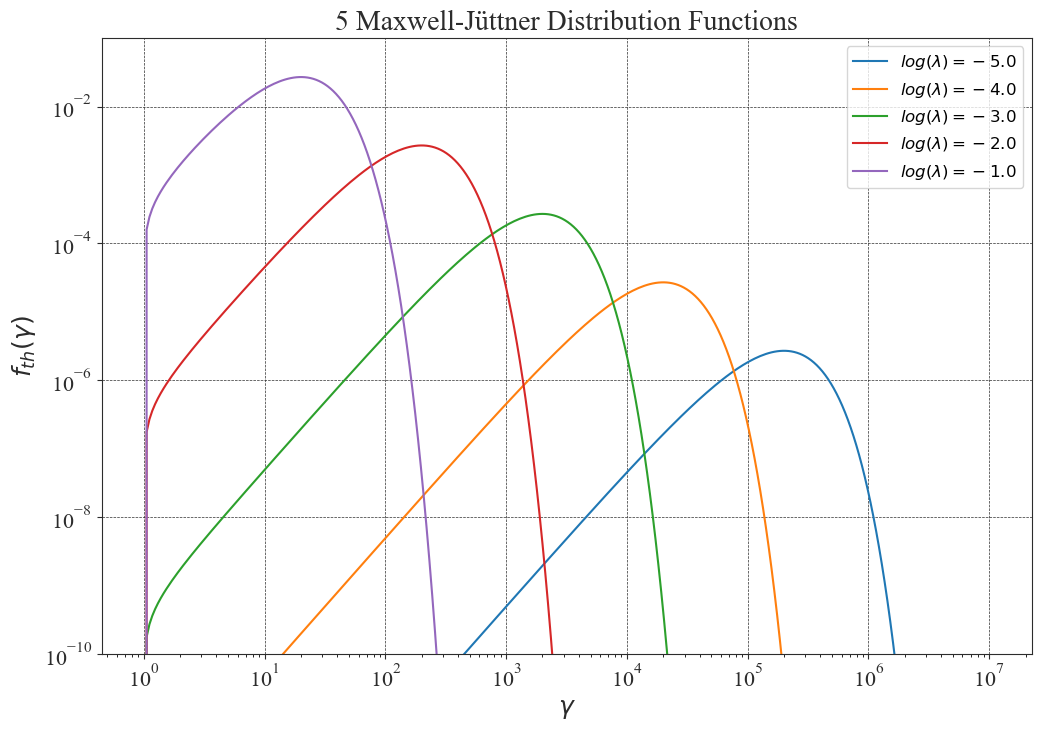}
    \caption{Five Maxwell-Jüttner distribution functions for a few different electron temperatures $\Theta_e = 1/\lambda$.}
    \label{fig:5MJ}
\end{figure}

A commonly used electron distribution function is the Maxwell-Jüttner distribution, a relativistic thermal distribution given by:
\begin{equation}
    f_{th} (\gamma, \Theta_e) = \frac{\gamma \sqrt{\gamma^2 - 1}}{\Theta_e K_2 (1/\Theta_e)}e^{-\gamma/\Theta_e}
    \label{maxwell jutt 2}
\end{equation}
where $\Theta_e \equiv \frac{k_B T_e}{m_e c^2}$ is the dimensionless electron temperature. Notably, for relativistic hot electrons $\Theta_e \gg 1$ the approximation $K_2 (1/\Theta_e) \approx 2\Theta_e^2$ holds. To be consistent with \citet{moscibrodzka2024stochastic} notation (see next subsection), in this paper we introduce $\lambda \equiv \Theta_e^{-1}$ to simplify both coding and notation, which results in the following expression:
\begin{equation}
    f_{th} (\gamma, \lambda) = \frac{1}{2} \gamma\sqrt{\gamma^2-1}\lambda^3 e^{-\gamma \lambda}.
    \label{maxwell jutt 2}
\end{equation}
As shown in Figure \ref{fig:5MJ}, this electron distribution function shows a singular peak with a strong cutoff, and as the temperature increases, the function shifts to the right. The Maxwell-Jüttner distribution function will serve as the basis function of the new method and will from this point on be referred as a `thermal component'.

\begin{figure}
    \centering
    \includegraphics[width=1\linewidth]{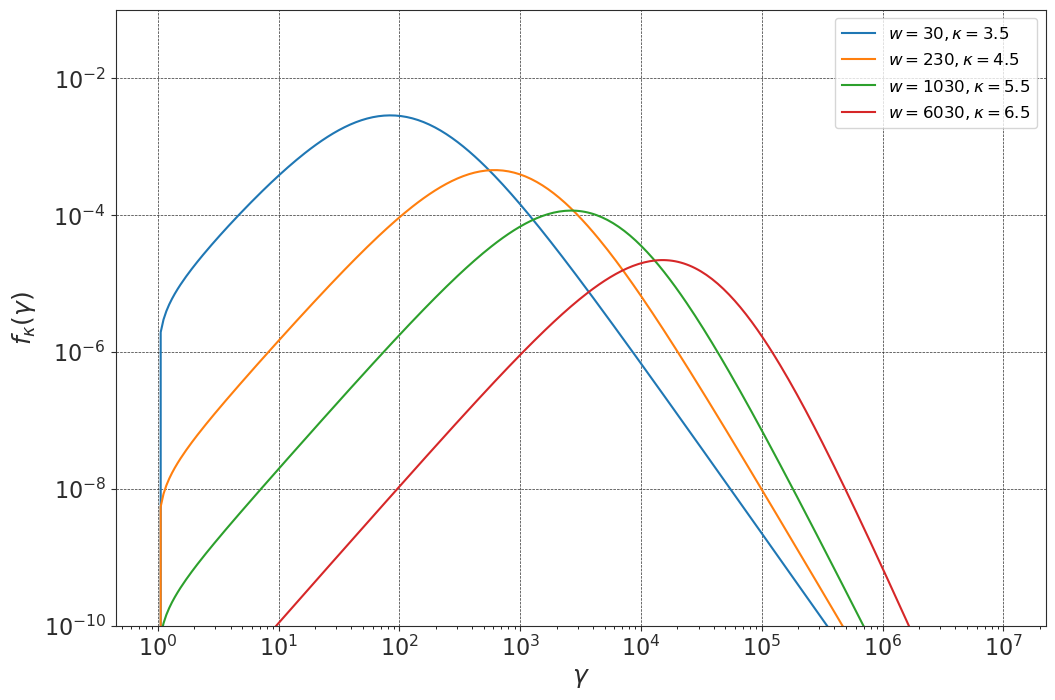}
    \caption{Four $\kappa$ distribution functions for a few different values of $w$ and $\kappa$ parameters.}
    \label{fig:4kappa}
\end{figure}

Another commonly used electron distribution function is the relativistic $\kappa$ distribution function, as proposed by \cite{xiao2006modelling}, which is given by the following:
\begin{equation}
    f_{\kappa} (\gamma) = N(\kappa,w) \gamma \sqrt{\gamma^2 -1} (1+\frac{\gamma-1}{\kappa w})^{- (\kappa+1)}
    \label{kappa dist 1}
\end{equation}
where $N(\kappa,w)$ is a normalization constant, $w$ is the parameter that influences the height of the peak of the function, as illustrated in Figure~\ref{fig:4kappa}, and $\kappa$ is a parameter that controls the steepness of the power-law tail of the function at higher energies. Also, notice that when $\kappa \rightarrow \infty$, $f_{\kappa}(\gamma)\rightarrow f_{th}(\gamma)$.

In this work, the condition $\kappa w \gg 1$ will generally hold, allowing for the simplification $N (\kappa, w) \approx  (\kappa-2) (\kappa-1)/2\kappa^2w^3$, which immediately implies $\kappa > 2$ and results in the following expression:
\begin{equation}
    f_{\kappa} (\gamma) = \frac{ (k-2) (k-1)}{2\kappa^2 w^3} \gamma \sqrt{\gamma^2 -1} (1+\frac{\gamma-1}{\kappa w})^{- (\kappa+1)}.
    \label{kappa dist 1}
\end{equation}
The relativistic $\kappa$ distribution function will be used as the trial function for the new method. Later in this work, multiple instances of the $\kappa$ function will be combined to add complexity, allowing to explore the new method's accuracy.

Currently, only a few electron distribution functions have analytical expressions for the transfer coefficients\footnote{The power-law distribution function is another common electron distribution function for which analytic transfer coefficients exist. This function will not be discussed in this work.}. Additionally, introducing new electron distribution functions requires the difficult and time-consuming evaluation of transfer coefficients, making it challenging to evaluate synchrotron transfer coefficients. 

\subsection{Stochastic averaging}

The method outlined in this article builds on the previous work of \cite{moscibrodzka2024stochastic}. 
In their work, any electron distribution function is approximated using the following (so called stochastic
average) equation:
\begin{equation}
    f_{avg} = \int d\lambda F (\lambda) f (\gamma, \lambda)
    \label{stoch avg}
\end{equation}
where $F(\lambda)$ represents a distribution of $\lambda$ (inverse of electron temperature) parameters and where $F(\lambda)$ is satisfying the normalization condition $\int d\lambda F(\lambda) = 1$. Following \citet{schwadron2010superposition} idea for non-relativistic electrons,  \citet{moscibrodzka2024stochastic} showed that the relativistic $\kappa$ distribution function can be well represented by a sum of relativistic thermal components and evaluated analytic expression for the $F(\lambda)$ function:
\begin{equation}
    F (\lambda) = \frac{1}{\Gamma (1-q)\lambda_0} e^{-\lambda/\lambda_0} (\frac{\lambda_0}{\lambda})^q
    \label{big F}
\end{equation}
where $q=3-\kappa$, $\lambda_0=1/w\kappa$, and $\Gamma$ stands for the complete Gamma function. 
The analytic expression has been evaluated only for the $\kappa$ distribution function, whereas the method developed here adopts a general numerical approach aimed at computing a wider range of electron distribution functions. Nevertheless, our numerical results will be compared to Equation~\ref{big F} as it constitutes a useful benchmark. 

\section{Method}\label{sec:method}

In this section, the methodology used to compute synchrotron transfer coefficients fast and accurate is outlined. The process begins by using the assumption that transfer coefficients have additive properties, and thus any distribution function can be described as a weighted sum. Then it progresses to finding an expression for the weights using the Integrated Squared Error function, and lastly solving that expression for the weights using Quadratic programming and the Clarabel model.

\subsection{The weighted sum approach}

Any arbitrary electron distribution function can be represented as a weighted sum of thermal components:
\begin{equation}
    f_{Chorus} (\gamma) = \sum^N_i w_i f_{th} (\gamma, \lambda_i)
    \label{weighted sum}
\end{equation}
where $f_{Chorus}$ is the approximation of any arbitrary electron distribution function, whether chosen or observed in the data. The $f_{th}$ are the basis thermal components and the $w_i$ are the weights that must be determined. Notice that Equations~\ref{stoch avg} and \ref{weighted sum} essentially represent the same process.

\begin{figure}
    \centering
    \includegraphics[width=1\linewidth]{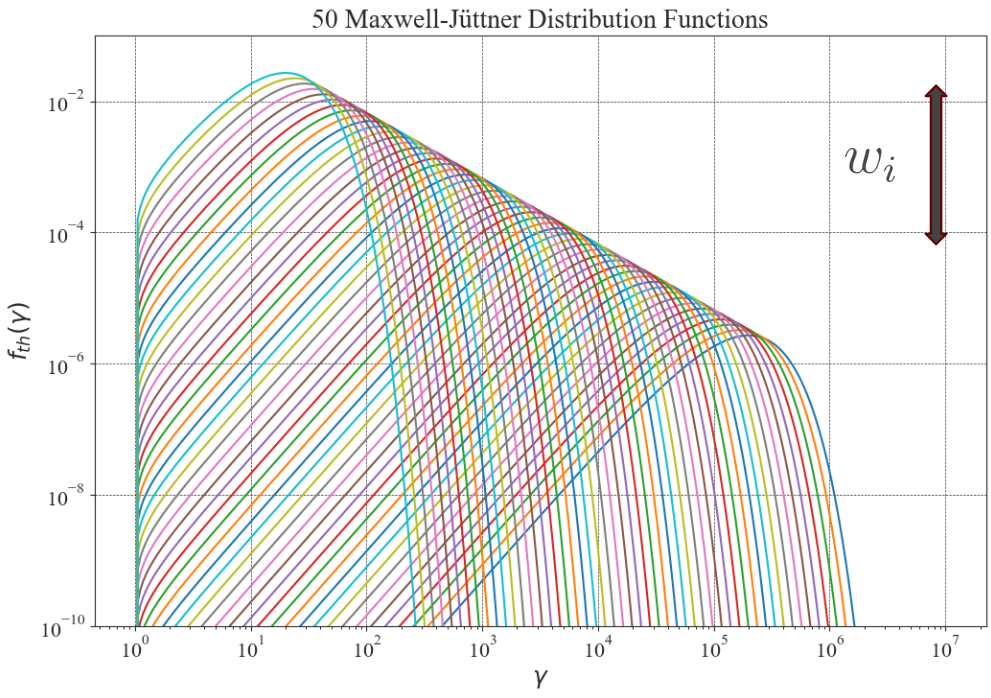}
    \caption{$N$=50 thermal components that each have a weight attached to adjust it either up or down.}
    \label{fig:50th}
\end{figure}

Figure~\ref{fig:50th} illustrates $N=50$ thermal components with different $\lambda$ (temperature) parameter where each thermal component is assigned a different weight. The task then becomes finding expression for the weights so that the resulting approximation aligns as closely as possible to the chosen or observed electron distribution function.

\subsection{Integrated Squared Error function}
To determine an expression for the weights, the Integrated Squared Error  (ISE) function is employed. This function integrates the squared difference between two functions over a chosen parameter. Alternatively, the Mean Integrated Squared Error  (MISE) is often used, which extends the ISE by computing the mean of the squared differences. Since the goal here is to derive an expression, either function would suffice. However, we use the Integrated Squared Error function for simplicity. The following is the general form of the Integrated Squared Error function:
\begin{equation}
    ISE = \int (f_{true} - f_{Chorus})^2 d\gamma 
    \label{ISE1}
\end{equation}
where $f_{true}$ represents any arbitrary electron distribution function, $f_{Chorus}$ is its approximation, and $\gamma$ is the chosen parameter for this method. The ISE function must then be minimized to reduce the error as much as possible. This minimization is achieved with respect to the weights, as they are the only unknown and adjustable variables in the equation, and is expressed as follows:
\begin{equation}
    ISE = \min_{w_i}\int (f_{true} - \sum^N_i w_i f_{th} (\gamma, \lambda_i))^2 d\gamma.
    \label{ISE2}
\end{equation}
Notice that the expression inside the integral takes the shape of a parabola, resulting in a single unique minimum. This is crucial because it eliminates the need for optimization methods, such as stochastic gradient descent, to locate the optimal minimum. The next step is to take the derivative of the minimization function with respect to the weights and set it equal to zero, which leads to the following equation:
\begin{equation}
    \frac{\partial  ISE}{\partial w_j} = -2\int (f_{true} - \sum^N_i w_i f_{th} (\gamma, \lambda_i))f_{th} (\gamma, \lambda_j) d\gamma = 0.
    \label{ISE3}
\end{equation}
After some simplifications, the expression simplifies to:
\begin{equation}
    \sum^N_i w_i \int f_{th} (\gamma, \lambda_i)f_{th} (\gamma, \lambda_j) d\gamma = \int f_{true}f_{th} (\gamma, \lambda_j) d\gamma.
    \label{simplified ISE}
\end{equation}
Next, we convert this expression into vectors and matrices, as this format is more suitable for computation in Python and simplifies the process of finding the weights, and yields the final formula:
\begin{equation}
    \vec f_{th} \vec f_{th}^T \vec w  = f_{true} \vec f_{th} \quad \rightarrow \quad A \vec w = \vec b.
    \label{simplified ISE2}
\end{equation}
Here, $A$ is an $N\times N$ matrix containing thermal components $\vec f_{th}$ multiplied by the transpose of the same set of thermal components $\vec f_{th}^T$. $\vec w$ represents the weights, and $\vec b$ is a vector formed by multiplying the chosen electron distribution function $f_{true}$ with the thermal components $\vec f_{th}$. Notice that the matrix $A$ is large and scales quadratically with the number of thermal components. This could potentially pose problems for computation time. However, it is important to recognize that $A$ depends entirely on the thermal components, which are known. As a result, matrix $A$ can be calculated only once and saved for different $N$ of thermal components. 

\subsection{Quadratic programming and the Clarabel model}

The next objective is to solve Equation \ref{simplified ISE2} for the weights $\vec w$. A straightforward approach might involve computing the inversion of matrix $A$ and solving the equation $\vec w = A^{-1}\vec b$. However, this method is unsuitable for two primary reasons.

First, matrix $A$ does not enforce non-negative weights. To facilitate physical interpretability and numerical stability of the implementation, it is better if the weights are constrained to be non-negative. Second, the elements of matrix $A$ are often close to zero, making $A$ nearly singular. As a result, an exact inverse of $A$ does not always exist. To address this, one might consider using the pseudo-inverse of $A$. However, in practice, this approach has shown to be inaccurate for more complex problems.
 
To solve Equation~\ref{simplified ISE2}, a method is required that ensures that the weights remain positive. A widely used approach in data science is Quadratic Programming, which solves equations involving a minimization function while enforcing constraints. The formulation for this method is as follows:
\begin{equation}
    \begin{aligned}
    \min_{w_i} \frac{1}{2}\vec w^T &A \vec w - \vec b^T\vec w\\
    &\vec w \geq 0
    \end{aligned}
    \label{minim QP}
\end{equation}
Appendix~\ref{app:A} provides additional context for this equation, emphasizing that matrix $A$ is symmetric and positive-definite. This property is always true for $A$ due to the way it is constructed.

\begin{figure*}
    \centering
    \includegraphics[width=0.95\linewidth]{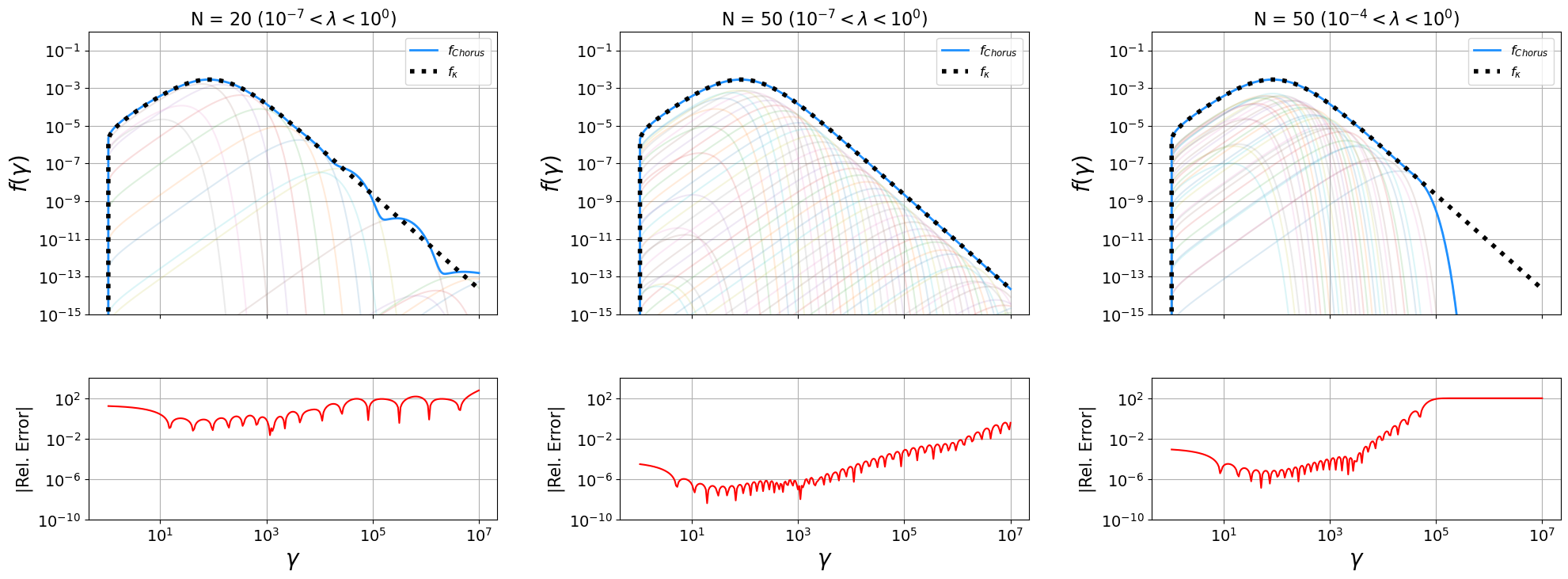}
    \caption{Chorus decomposition (shown as solid line) of the $\kappa$ distribution function (dotted line) into $N=20$ (left panel) and $N=50$ (middle and right panels) thermal components. The parameters of the $\kappa$ function are $\kappa = 3.5$ and $w = 30$. Parameter $\lambda$ ranges from $10^{-7}$ to $10^{0}$ in the left and middle panels and from $10^{-4}$ to $10^{0}$ in the right panel. The error shown under each panel is computed according to Equation~\ref{error1}. }
    \label{fig:kapp aprox}
\end{figure*}
 
There are various approaches to solving Equation~\ref{minim QP}. One method incorporates the restriction directly into the minimization function as an additional term that penalizes negative weights, for example, using Lagrange multipliers or min-max functions. However, this work focuses on a numerical approach using interior point solvers\footnote{The preference for an interior point solver over Lagrange multipliers stems from the fact that interior point methods handle inequality constraints more naturally and scale better with a larger number of variables.}, a Python library developed by \cite{goulart2024clarabel}.
One such solver is Clarabel. The approach of this solver in terms of Chorus variables is as follows, define the optimization problem:
\begin{equation}
    \begin{aligned}
    \min_{\vec w, \vec s} &\frac{1}{2}\vec w^T A \vec w - \vec b\vec w\\
     \text{subject to} \quad &L\vec w+ \vec s = \vec t\\
     &\vec s \in \mathbb{K}
    \end{aligned}
    \label{clarabel 1}
\end{equation}

Here, $A \in \mathbb{R}^{n \times n}$, $L \in \mathbb{R}^{m \times n}$, with decision variables $w \in \mathbb{R}^{n}$, $s \in \mathbb{R}^{m}$. Additionally, there are vectors $t \in \mathbb{R}^{m}$ and $q \in \mathbb{R}^{n}$. The set $\mathbb{K}$ is a set of a closed and convex cone, which can be seen has hard boundaries for the primal function. Equation \ref{clarabel 1} is commonly referred to as the 'primal' function, and it coincides with the 'dual' function, which is defined as follows:

\begin{equation}
    \begin{aligned}
    \max_{\vec w, \vec z} -&\frac{1}{2}\vec w^T A \vec w - \vec t^T\vec w\\
     \text{subject to} \quad &A\vec w + L^T\vec z = \vec b^T\\
     &\vec z \in \mathbb{K^*}
    \end{aligned}
    \label{clarabel 2}
\end{equation}
Here, the set $\mathbb{K^*}$ is called the dual cone, representing the hard boundary for the dual function. The primal and dual functions are strongly related but offer different approaches to solving an optimization problem. These functions also allow us to define the duality gap as follows:

\begin{equation}
    \begin{aligned}
    \delta &=  (\frac{1}{2}\vec w^T A \vec w - \vec b\vec w)- (-\frac{1}{2}\vec w^T A \vec w - \vec b^T\vec w)\\
     &= \vec w^T A \vec w - \vec b \vec x + \vec t^T \vec z = \vec s^T \vec z.
    \end{aligned}
    \label{clarabel 2}
\end{equation}
The final substitution stems from the KKT (Karush-Kuhn-Tucker) conditions outlined in Equation \ref{QP appendix 1}, which were first introduced and explained by \cite{kuhn1951nonlinear}. The duality gap is central to this optimization method, as minimizing this gap yields the best results for the decision variables. The Clarabel Python library can accept an initial guess for the decision variables, although setting them to zero is still acceptable. Afterward, the code iteratively adjusts the decision variables until the duality gap function reaches zero. There are additional processes occurring behind the scenes, so it is highly recommended to read the paper on Clarabel by \cite{goulart2024clarabel} for further clarification and additional details.

As a side note, if the $\vec w^T A \vec w$ term is set to zero, this method would be classified as linear programming. In general, linear programming is much faster than quadratic programming, but it is significantly less accurate. Given that this project involves fairly complicated problems, linear programming quickly becomes inaccurate. The reason for this is that quadratic programming contains quadratic terms in $\vec w$, which causes the iterative process in Clarabel to take smaller steps. While this results in a longer time to reach the optimal weights, it makes the method less prone to overfitting. Additionally, the off-diagonal elements in matrix $A$ capture interactions between thermal components, providing additional information that linear programming does not utilize.
 
Here, the restriction $\vec w \geq 0$ is formulated by setting matrix $L$ as the identity matrix $I_n$ and vector $\vec t = \vec 0$, while the nonnegative cone is chosen to enforce the inequality constraint, meaning $\vec s, \vec z \in \mathbb{R}^n_{\geq0}$.

When accurate weights are determined using this method, the same weights can be directly applied to the transfer coefficients. E.g. for emissivity, this can be expressed as follows:
\begin{equation}
    j_{\nu} (f_{Chorus}) = j_{\nu} (\sum^N_i w_i^{*} f_{th,i}) = \sum^N_i w_i^{*} j_{\nu,i} (f_{th,i}).
    \label{final emiss example}
\end{equation}
Here, $w_i^*$ represents the normalized weights, ensuring that $\sum^N_i w_i^* = 1$. Normalizing the weights is essential for making a fair comparison with another model, in this case Symphony, which will be shown later in the results.
 
All code and data used in Chorus is available online through Github. The repository contains a Jupyter Notebook and all .csv data used.\footnote{Github: https://github.com/DAvanDuren/Chorus.}

\section{Results}

\subsection{Decomposition of electron distribution function}

\subsubsection{$\kappa$ distribution function}\label{sec:kappa_decompose}

\begin{figure*}
    \centering
    \includegraphics[width=1\linewidth]{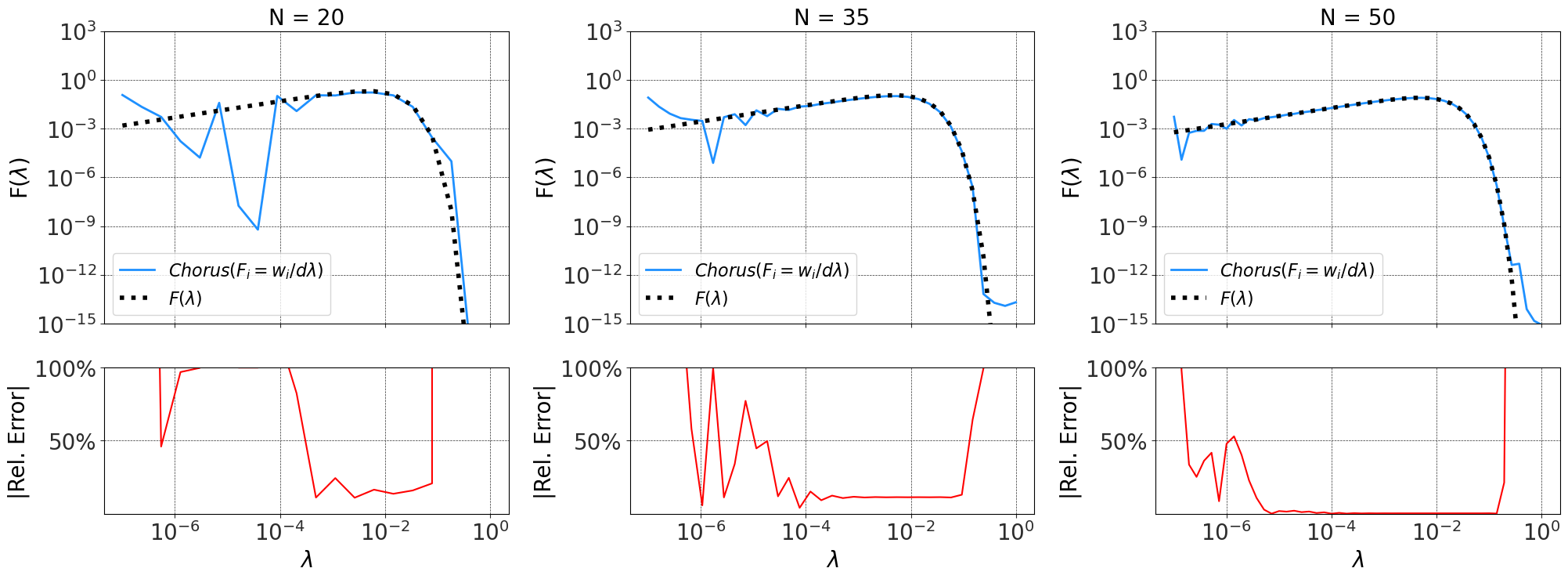}
    \caption{Comparison between analytic $F(\lambda)$ from Equation~\ref{big F} \citep{moscibrodzka2024stochastic} and the weights computed by the our numerical method divided by $d\lambda$ for $N = 20, 35, 50$ components. Additionally we show difference between analytic and numerical results. The parameters used: $\kappa = 3.5$, $w = 30$ and $\lambda$ from $10^{-7}$ to $10^0$. The relative error is calculated using Equation~\ref{error1}.}
    \label{fig:compare stoch avg}
\end{figure*}

Using the method described in Section~\ref{sec:method}, the first result is the decomposition of the $\kappa$ distribution function (Equation~$\ref{kappa dist 1}$) into a sum of thermal components (Equation~\ref{maxwell jutt 2}). Figure~\ref{fig:kapp aprox} shows the Chorus code result for two values of $N$ and two ranges of $\lambda$ parameter together with the relative error calculated using the following error formula:
\begin{equation}
    |Rel. Error| = \frac{|f_{Chorus} - f_{\kappa}|}{f_{\kappa}} \cdot 100\%.
    \label{error1}
\end{equation}

\begin{table}
\centering
\begin{tabular}{lccc}
\hline
 $N$ & $20$ & $50$  & $50$  \\
  $\lambda_{\rm min}$ & $10^{-7}$ & $10^{-7}$ & $10^{-4}$ \\
\hline
 Computing time & 0.71 ms & 6.29 ms &  6.16 ms\\
 Max. error &  599\%&  0.39\%&  100\%\\
 Median error &  $7.75$\% &  $8.48 \times 10^{-6}$\% & $7.44 \times 10^{-4}$\% \\
 Min. error &  0.22\% &  $4.04\times 10^{-9}$\% &  $1.33\times10^{-7}$\%\\
\hline
\end{tabular}
\caption{Computing times for the decomposition and errors between Chorus and analytic expression as a function of number of components and $\lambda$ span for the $\kappa$ distribution function.}\label{table:1}
\end{table}

In Table~\ref{table:1}, we report the computational time for each Chorus decomposition as a function of $N$ and lower limit for the $\lambda$ parameter, $\lambda_{\rm min}$. The computational times are given excluding the computation time for the matrix $A$. For the middle case ($N=50$ and $\lambda_{\rm min}=10^{-7}$) the errors are minimal. In two cases, we see a steep increase in error at higher energies, one due to poor reconstruction and the second one due to limited range of $\lambda$, which produces a cut-off at high energies. 

Theoretically, function $F(\lambda)$ from Equation~\ref{big F} should correspond to the weights derived numerically divided by $d\lambda$. Figure~\ref{fig:compare stoch avg} presents a comparison between analytic $F(\lambda)$ and numerically evaluated $w_i/d\lambda$ for three different values of $N$. It is evident that our numerical solution for $w_i$ converges to the analytic expression when increasing $N$. 
The median difference between the two approaches is $0.8\%$ for $N = 50$. Assuming setting $N>50$ does not further decrease the error. The accuracy of the decomposition could be further improved by increasing the number of weights along with expanding the range of temperatures used in the calculation.

\subsubsection{Complicated distribution function}\label{sec:complexDF}

Since our numerical method is general, it can handle functions more complicated than the $\kappa$ function. To simulate a more composite function, three $\kappa$ distribution functions with different parameters are summed:
\begin{equation}
\begin{aligned}
f(\gamma) \equiv f_{\kappa 1}(\kappa=3.5,w=10)+ \\  f_{\kappa 2}(\kappa=6,w=1000)+ \\ f_{\kappa 3}(\kappa=4,w=2000).
\end{aligned}
\end{equation}

\begin{figure*}
    \centering
    \includegraphics[width=0.95\linewidth]{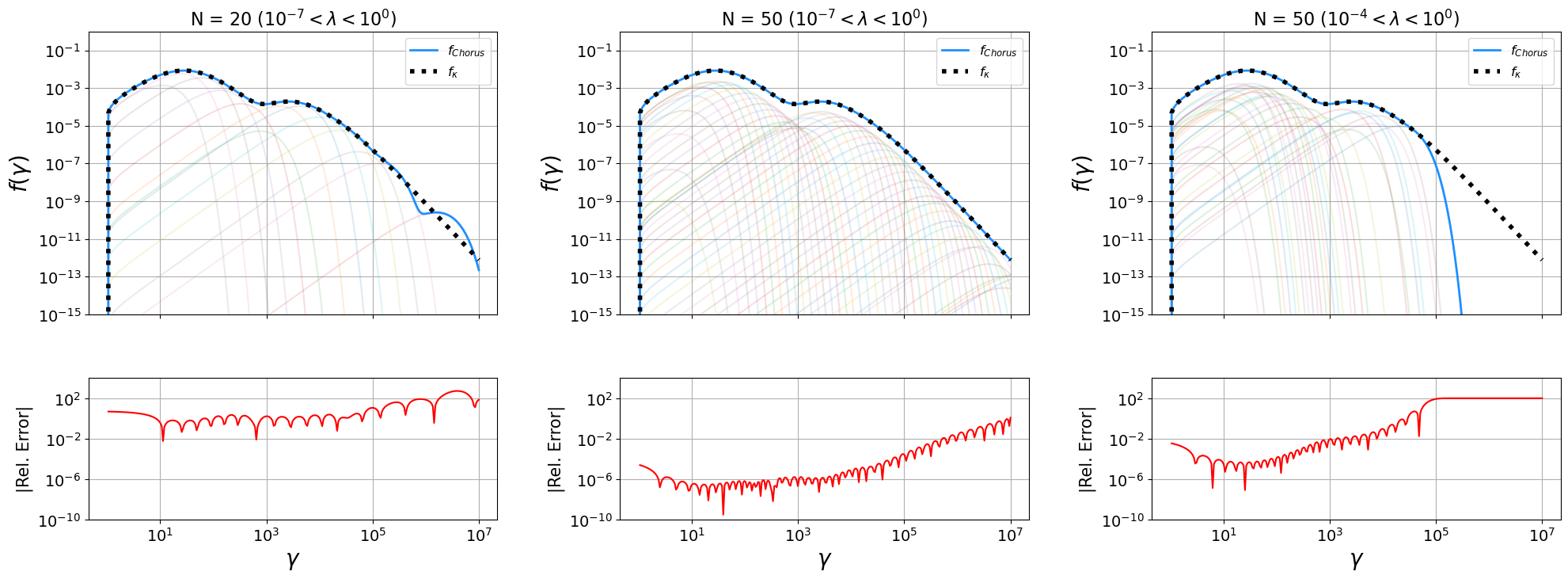}
    \caption{Same as Figure~\ref{fig:kapp aprox} but for the complicated function (sum of three distinct $\kappa$ distribution functions).}
    \label{fig:3kappa approx}
\end{figure*}

\begin{table}
\centering
\begin{tabular}{lccc}
\hline
 $N$ & $20$ & $50$  & $50$  \\
 $\lambda_{\rm min}$ & $10^{-7}$ & $10^{-7}$ & $10^{-4}$ \\
\hline
 Computing time & 1.44 ms & 5.97 ms &  7.05 ms\\
 Max. error & 529\%&  1.19\%&  100\%\\
 Median error & 1.78\%& $1.88\times 10^{-6}$\%& $1.30\times 10^{-2}$\%\\
 Min. error & $5.74\times 10^{-3}$\%& $2.98\times10^{-10}$\%&  $8.06\times 10^{-8}$\%\\
\hline
\end{tabular}
\caption{Sames as Table~\ref{table:1} but for the complicated distribution function.}\label{table:2}
\end{table}

Figure~\ref{fig:3kappa approx} shows the decomposition of the complicated function into a sum of thermal components for two values of the parameter $N$ and two different ranges of $\lambda$. Table~\ref{table:2} reports the computational times and errors between the numerical and analytic results. The results obtained for the complicated distribution function are comparable to those for the single $\kappa$ distribution function. 

\subsection{Synchrotron transfer coefficients}

In what follows, we present a comparison of selected synchrotron transfer coefficients calculated using our Chorus scheme with those directly integrated with the Symphony code. In these calculations, we assume a default pitch angle $\theta = 60$ degrees and magnetic field strength $B = 30$ G for which the cyclotron frequency is $\nu_c = eB / (2 \pi m_e c) = 0.84 \times 10^8$ Hz. All transfer coefficients are number density normalized.

\subsubsection{$\kappa$ distribution function}

\begin{figure*}
    \centering
    \includegraphics[width=0.95\linewidth]{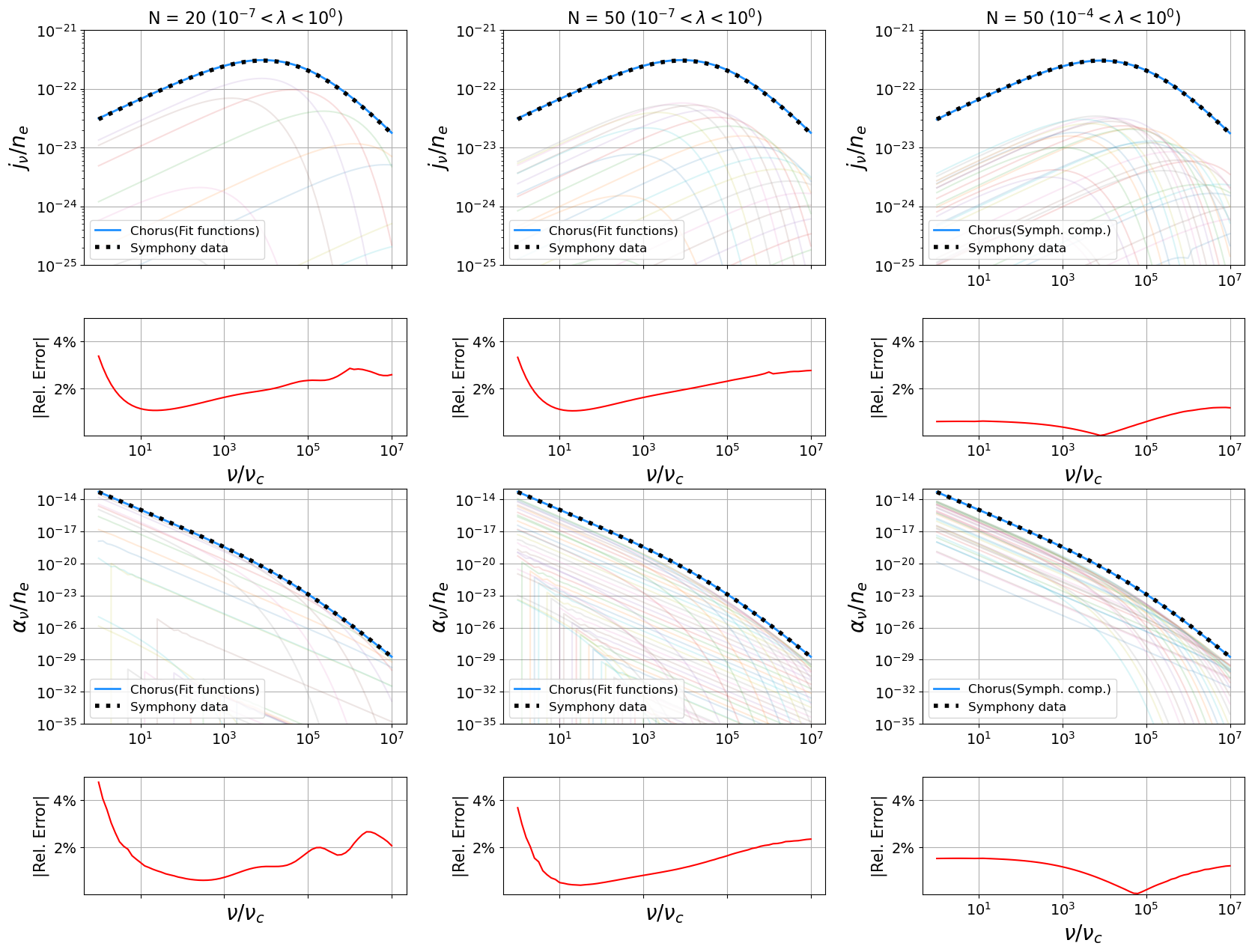}
    \caption{Density-normalized emissivity (top panels) and absorptivity (lower panels) of the $\kappa$ distribution function build-up using $N=20$ (left panel) and $N=50$ (middle panel) thermal components approximated with fit functions of \citet{pandya2016polarized} (see also \citealt{leung2011numerical} and Appendix~\ref{app:B})
    compared to the direct numerical integration of the $\kappa$ emissivity/absorptivity using Symphony. All thermal components are shown as lightly transparent color lines. The right panels show the same comparison of emissivity and absorptivity but when they are build-up using $N=50$ thermal components where also each thermal component is directly integrated with Symphony instead of being a fit function. Notice, that in the right panels the range of thermal components is smaller compared to the left and middle panels.}\label{fig:all good}
\end{figure*}

Figure~\ref{fig:all good} shows the performance of the Chorus scheme for emissivity and absorptivity in Stokes $I$ for the $\kappa$ distribution function from Section~\ref{sec:kappa_decompose}. In the left and middle panels the emissivity and absorptivity thermal sub-components are the thermal fit functions from \citet{pandya2016polarized} (see also \citealt{leung2011numerical} and Appendix~\ref{app:B}) assuming $N=20$ and $N=50$ sub-components. In the right panel, again $N=50$ sub-components is used but the thermal sub-components are not fit functions but thermal emissivities and absorptivities integrated directly in Symphony. Notice that in the latter case the range
of $\lambda$ decreases.
In Table~\ref{table:3} we report errors in recovered emissivities and absroptivities
with respect to direct integration with Symphony. 
In all cases, these errors are at a level of approximately 2\% for most frequencies. Substituting thermal fit functions with numerically integrated thermal emissivities decreases the reconstruction error of the method, but not significantly. We conclude that
neither larger number of thermal subcomponents nor more accurate thermal sub-components increase the accuracy of emissivities and absorptivities derived in our scheme. The computational times of Chorus reported in Table~\ref{table:3} are short, on the order of a few ms. For comparison, Symphony integrates $\kappa$ emissivity and absorptivity in 10 to 20 minutes per transfer coefficient depending on the variables involved.

\begin{table}
\centering
\begin{tabular}{lccc}
\hline
 $N$ & $20$ & $50$  & $50$  \\
thermal component & fit func. & fit func. & Symphony\\
 \hline
\multicolumn{4}{c}{ $j_{\rm I}$}\\
\hline
$\lambda_{\rm min}$ & $10^{-7}$ & $10^{-7}$ & $10^{-4}$ \\
 Computing time & 1.37 ms & 7.42 ms &  53.7 ms\\
 Max. error & 3.38\%&  3.33\%&  1.53\%\\
 Median error & 1.91\%& 1.93\%& 0.59\%\\
 Min. error & 1.08\%& 1.05\%& 0.04\%\\
\hline
\multicolumn{4}{c}{$\alpha_{\rm I}$}\\
\hline
$\lambda_{\rm min}$ & $10^{-7}$ & $10^{-7}$ & $10^{-4}$ \\
 Computing time & 1.56 ms & 7.86 ms &  51.6 ms\\
 Max. error & 4.76\%&  3.68\%&  1.53\%\\
 Median error & 1.31\%& 1.19\%& 1.12\%\\
 Min. error & 0.60\%& 0.39\%& 0.04\%\\
\hline
\multicolumn{4}{c}{ $\rho_{\rm Q}$}\\
\hline
$\lambda_{\rm min}$ & $10^{-7}$ & $10^{-4}$ & - \\
computing time & 2.08 ms & 9.12 ms &  -\\
Max. error & 49.3\%&  83.7\%&  -\\
Median error & 7.45\%& 8.18\%& -\\
Min. error & 0.27\%& 0.13\%& -\\
\hline
\end{tabular}
\caption{Computing times and errors of Chorus when estimating transfer coefficients for $\kappa$ distribution function. All computing times include the computation time for the weights. Notice that computing times reported in the last column are increased mainly due to reading off the numerical data from files.}\label{table:3}
\end{table}

\begin{figure*}
    \centering
    \includegraphics[width=1\linewidth]{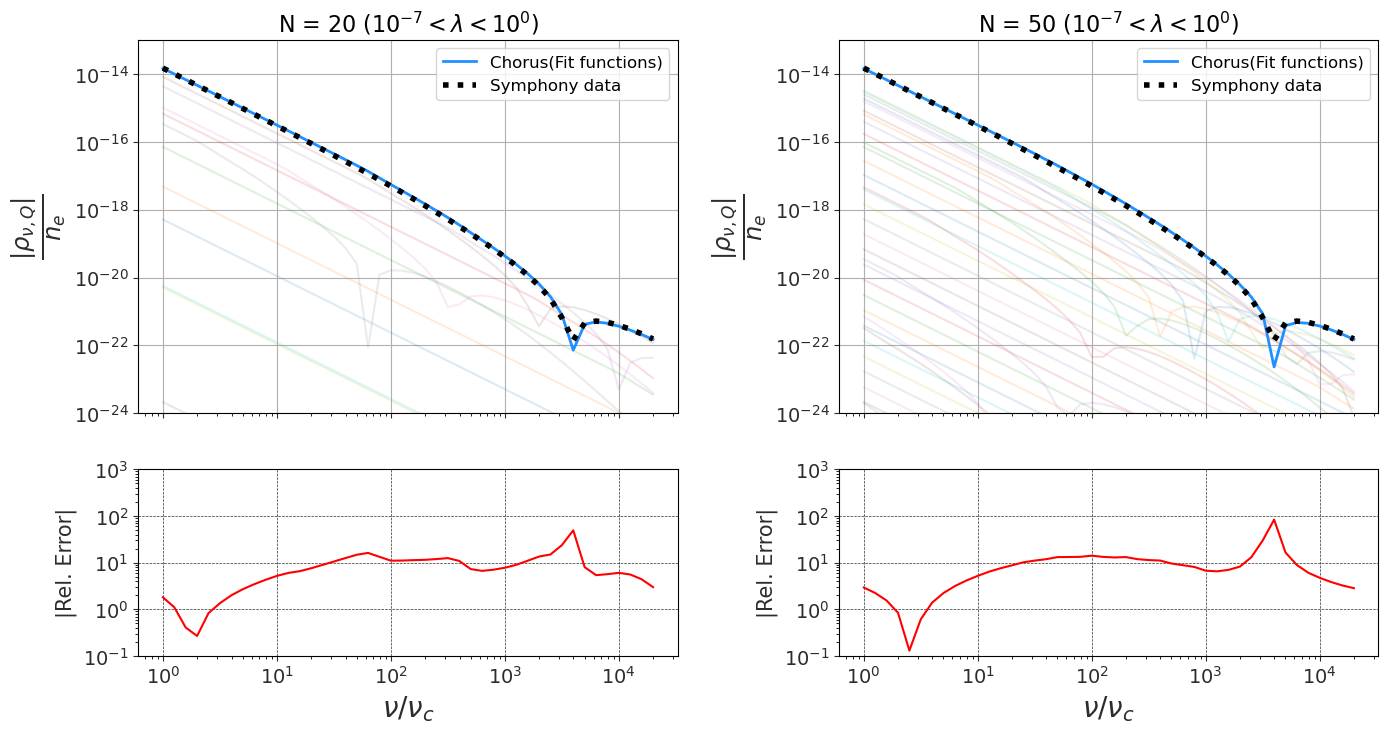}
    \caption{Same as in Figure~\ref{fig:all good} but for density-normalized Faraday conversion coefficient $\rho_Q$.}
    \label{fig:rotat approx}
\end{figure*}

Figure~\ref{fig:rotat approx} shows the performance of the Chorus scheme for the synchrotron rotativity in Stokes parameter Q (i.e. Faraday conversion) and compares it to direct integration of Symphony. Here, the calculation is performed again for two values of $N$, a single range of $\lambda$ and is limited to the use of the fit functions (see the formula in Appendix~\ref{app:B}).
Direct integration of rotativities with Symphony, even for the thermal distribution function, typically requires significantly more time compared to our weighted-average scheme (from days to weeks depending on plasma parameters). Because of this limitation, there are two additional differences in computing the results for the rotativity compared to the emissivity and absorptivity coefficients. For $\rho_{\rm Q}$, only 45 data points were considered, whereas the other two transfer coefficients used 71. Furthermore, the $\rho_{\rm Q}$ calculation extends only up to $\nu/\nu_c = 10^{4.4}$, as opposed to $10^7$ for the emission/absorption coefficients. This limitation is primarily due to the fact that Symphony’s computational time increases exponentially at higher values of $\nu/\nu_c$. Since this result serves as a demonstration of the new method, this range is deemed sufficient. Table~\ref{table:3} also reports computing times and errors for results shown in Figure~\ref{fig:rotat approx}. 

\subsubsection{Complicated distribution function}

\begin{figure*}
    \centering
    \includegraphics[width=0.95\linewidth]{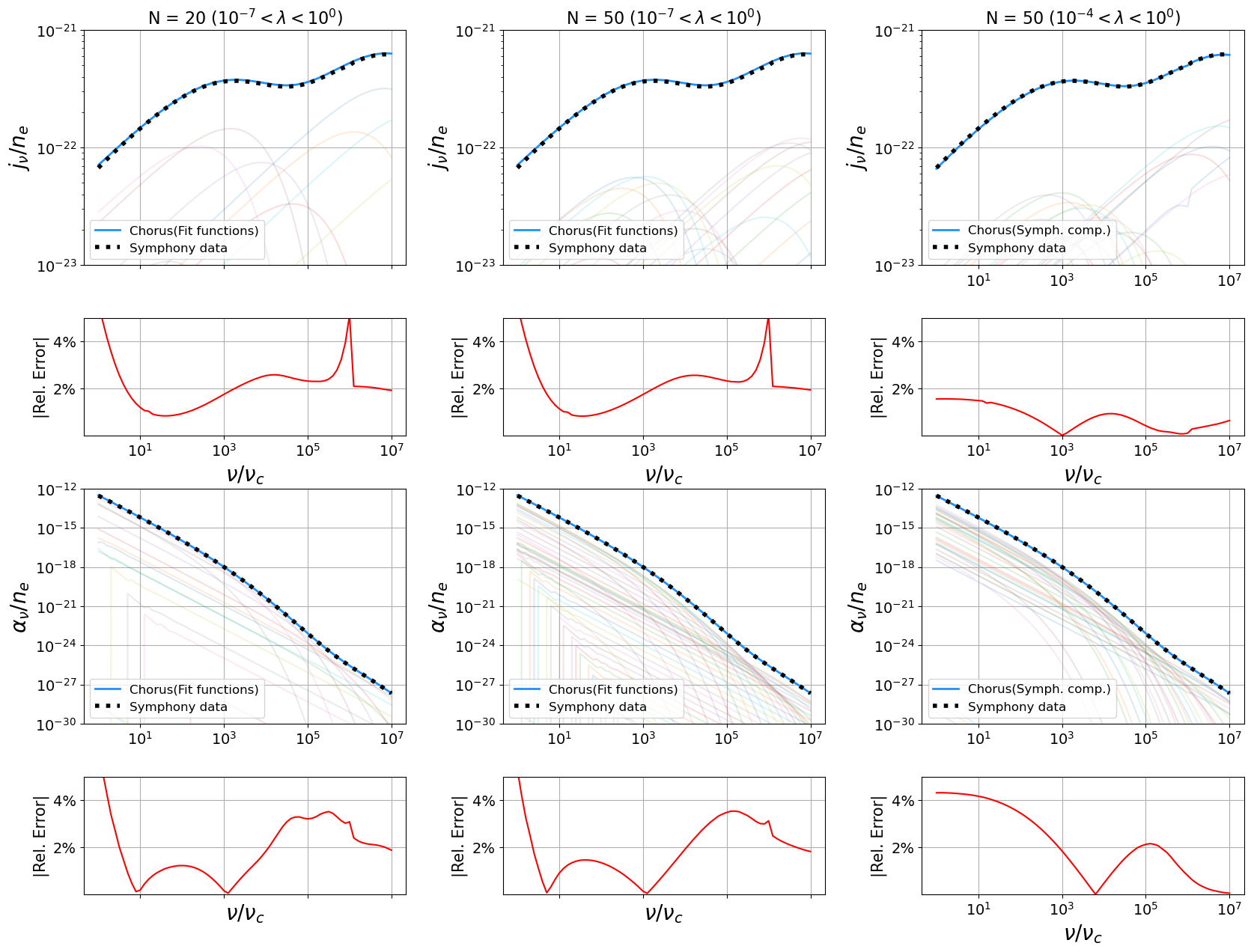}
    \caption{Same as in Figure~\ref{fig:all good} but for our complicated distribution function.}
    \label{fig:3kappa emiss and abs approx}
\end{figure*}

\begin{table}
\centering
\begin{tabular}{lccc}
\hline
 $N$ & $20$ & $50$  & $50$  \\
thermal component & fit func. & fit func. & Symphony\\
 \hline
\multicolumn{4}{c}{ $j_{\rm I}$}\\
\hline
$\lambda_{\rm min}$ & $10^{-7}$ & $10^{-7}$ & $10^{-4}$ \\
 computing time & 2.27 ms & 7.60 ms &  53.5 ms\\
 Max. error & 5.70\%&  5.59\%&  1.56\%\\
 Median error & 2.09\%& 2.08\%& 0.67\%\\
 Min. error & 0.84\%& 0.83\%& 0.02\%\\
  \hline
\multicolumn{4}{c}{ $\alpha_{\rm I}$}\\
\hline
$\lambda_{\rm min}$ & $10^{-7}$ & $10^{-7}$ & $10^{-4}$ \\
 computing time & 2.45 ms & 8.06 ms &  52.4 ms\\
 Max. error & 6.28\%& 5.29\%&  4.32\%\\
 Median error & 1.88\%& 1.73\%& 1.95\%\\
 Min. error & 0.05\%& 0.04\%& 0.01\%\\
\hline
\end{tabular}
\caption{Same as Table~\ref{table:3} but for complicated distribution function.}\label{table:4}
\end{table}

Figure~\ref{fig:3kappa emiss and abs approx} shows the synchrotron emissivity and absorptivity in Stokes $I$ for the complicated distribution function presented in Section~\ref{sec:complexDF}. Table~\ref{table:4} presents errors and computing times. On average, the errors are below 2\% for both coefficients. In this case again neither more thermal sub-components nor more accurate thermal sub-components lead to a better emissivity/absorptivity approximation. 

\section{Discussion}

This paper presents a new method for approximating synchrotron transfer coefficients for arbitrary electron distribution functions. 

While the developed method demonstrates significant computational efficiency and accuracy in approximating synchrotron transfer coefficients for any arbitrary electron distribution functions, it is essential to address its limitations and the broader implications of its use. These limitations highlight the challenges the method may face, while the implications shed light on its potential applications and future refinements.

A key limitation of the method is the significant increase in error at higher energies when approximating the electron distribution function, as seen in, e.g., Figure~\ref{fig:kapp aprox}. This discrepancy likely stems from the inherent limitations of Python's numerical precision. The approximations span $f (\gamma)$ values from approximately $10^{-3}$ to $10^{-15}$, a range covering 12 orders of magnitude. However, common Python libraries such as NumPy and SciPy typically use 64-bit floating-point precision, which provides a maximum precision of 15 to 17 significant digits.
As $f (\gamma)$ decreases at higher energies, the precision required to accurately adjust the weights diminishes. Limited numerical precision disproportionally affects smaller values, resulting in a less accurate approximation in these regions.
One could argue that this issue is not particularly significant, as $f (\gamma)$ values as small as $10^{-15}$ have a negligible impact on the transfer coefficients compared to values closer to $10^{-3}$. However, if this precision loss poses a problem for the user, several adjustments can be made. For instance, if the user's interest lies specifically in higher energies or frequencies, the range of $\gamma$ and $\lambda$ can be narrowed to focus only on the desired energy or frequency spectrum. This adjustment can reduce the difference between the maximum and minimum values $f (\gamma)$, which mitigates the problem of precision.
A more robust solution would involve implementing our method in a programming language which supports higher floating-point precision. Julia, for example, offers 128-bit floating-point precision, enabling approximately 20 to 23 significant digits of accuracy. Although this change might lead to an increase in computational time, it presents a viable alternative for cases where accuracy outweighs efficiency as the primary goal.

Another significant limitation of our method is the reliance on fit functions for calculating the transfer coefficients. Although both the emissivity and absorptivity for both demonstrations appear to be relatively accurate, the rotativity introduces substantial errors. These errors may arise because the method applies the fit functions $N$ times, with each fit function carrying its own intrinsic error. \cite{marszewski2021updated} demonstrates that the fit function for $\rho_{\rm Q}$ can introduce errors on the order of a few percent, and when this function is used $N$ times, the error can scale dramatically. Figure~\ref{fig:rotat approx} also shows that using fewer thermal components can improve the accuracy of the approximation, but this comes at the cost of reduced accuracy for the electron distribution function, as shown in Figure~\ref{fig:kapp aprox}. A possible improvement to the code would be to allow the user to optimize the number of thermal components according to the user's needs, though this may significantly increase runtime. A more comprehensive solution would involve developing better fit functions for the transfer coefficients, which are not yet available. As shown in the results, sub-components for Chorus can also be created by Symphony. When calculating emissivity and absorptivity, using Symphony sub-components slightly decreases errors but increases runtime, as N files need to be read by the code. Additionally, if a different number of sub-components or different parameter ranges are required, new sub-components must be generated using Symphony.

As a final remark, \citet{davelaar2024mlody} introduces MLody, a deep neural network designed to generate polarized synchrotron coefficients with high accuracy, specifically for black hole accretion studies. Trained on data from Symphony, MLody outperforms traditional fit functions across diverse plasma conditions, significantly improving polarization modeling for thermal Maxwellian electron distributions. Future developments will expand its applicability to more complicated electron distributions, including anisotropic distribution functions. Mlody offers a complementary solution to the method developed in this paper by avoiding the use of fit functions and its intrinsic inaccuracies. In contrast, Chorus is more economical and achieves much better runtimes.





\section*{Acknowledgements}

The authors thank Marijke Haverkorn for her comments on the work.

\section*{Data Availability}

The data underlying this article will be shared on reasonable request to the corresponding author.







\appendix

\section{Formulas for Quadratic programming}\label{app:A}

One of the origins of Equation \ref{minim QP} is the linear squares function, but the general form of quadratic programming is first discussed in the book named Numerical Optimization by \cite{nocedal2006quadratic} without a clear derivation. 
\begin{equation}
    \begin{aligned}
    \min_{w_i} ||R\vec x - \vec d||^2 &= \min_{w_i}  (R\vec x - \vec d) (R^T\vec x^T - \vec d^T)\\
    &= \min_{w_i} \vec x R R^T \vec x - 2R \vec d^T \vec x - \vec d\vec d^T\\
    &= \min_{w_i} \vec x^TQ\vec x + 2\vec c^T \vec x\\
    &= \min_{w_i} \frac{1}{2}\vec x^T Q \vec x + \vec c^T \vec x .
    \end{aligned}
    \label{QP appendix 1}
\end{equation}
Here, $Q = R^TR$ is called the 'Cholesky Decomposition', $\vec  c = R^T \vec d$ and $||\cdot ||^2$ is called the 'square of a norm'. Also, notice that $\vec d \vec d^T$ vanishes because the function is minimized with respect to $\vec x$ and $\vec d \vec d^T$ has no dependence on $\vec x$.
 
The following expressions are the KKT conditions:
\begin{equation}
    \begin{aligned}
    L\vec x + \vec s &= \vec t\\
    P \vec x + L^T\vec z &= -\vec q\\
    \vec s^T \vec z &= 0\\
     (\vec s, \vec z) &\in \mathbb{K} \times \mathbb{K^*}.
    \end{aligned}
    \label{QP appendix 2}
\end{equation}

\section{Fit functions for synchrotron transfer coefficients}\label{app:B}

\subsection{Thermal emissivity and absorptivity fit functions}

Fit functions to thermal synchrotron emissivity are provided by \citet{pandya2016polarized} (see also \citealt{leung2011numerical}) in the following form:
\begin{equation}
    j_S = \frac{n_e e^2 \nu_c}{c}J_S (\frac{\nu}{\nu_c}, \theta)
    \label{emiss leung}
\end{equation}
where $J_S$ is the dimensionless emissivity and $S$ stands for a Stokes parameter. A similar formula is introduced for absorptivity:
\begin{equation}
    \alpha_S = \frac{n_e e^2}{\nu m_e c} A_S (\frac{\nu}{\nu_c}, \theta)
    \label{absorp leung}
\end{equation}
where $A_S$ is the dimensionless absorptivity. Both $J_S$ and $A_S$ depend on angle $\theta$, frequency $\nu$, and electron cyclotron frequency $\nu_c=eB/(2\pi m_e c)=2.8 \times 10^6 B$~Hz. 

The dimensionless emissivity is:
\begin{equation}\begin{aligned}
    J_S =e^{-X^{1/3}} \times \\
    \begin{cases}
        \frac{\sqrt{2\pi}}{27} \sin (\theta) \left (X^{1/2} + 2^{11/12} X^{1/6} \right)^2, & \text{Stokes } I, \\[10pt]
        -\frac{\sqrt{2\pi}}{27} \sin (\theta) \left (X^{1/2} + \frac{7 \Theta_e^{24/25} + 35}{10 \Theta_e^{24/25} + 75} \, 2^{11/12} X^{1/6} \right)^2, & \text{Stokes } Q, \\[10pt]
        0, & \text{Stokes } U, \\[10pt]
        -\frac{37 - 87 \sin\left (\theta - \frac{28}{25}\right)}{100  (\Theta_e + 1)} 
        \left ( 1 + \left (\frac{\Theta_e^{3/5}}{25} + \frac{7}{10} \right) X^{9/25} \right)^{5/3}, & \text{Stokes } V
    \end{cases}
    \label{th dimen emiss}
    \end{aligned}
\end{equation}
where $X = \nu / \nu_s$ and $\nu_s=(2/9) \nu_c \Theta_e^2 \sin \theta$.
 
Based on Kirchoff's law, that reads $J_S - \alpha_S B_{\nu} = 0$, where 
\begin{equation}
    B_{\nu} = \frac{2h\nu^3}{c^2}\frac{1}{e^{h \nu /kT_e} -1}
    \label{planck function}
\end{equation}
is the Planck function, one can write a formula for the  dimensionless absorptivity as:
\begin{equation}
    A_S = \frac{J_S}{B_{\nu}} = J_S \frac{m_e c^2 \nu_c}{2h\nu^2} (e^{h\nu/kT_e}-1).
    \label{th dimen absorp}
\end{equation}

\subsection{Thermal rotativity fit}

An analytic approximation of synchrotron rotativity (Faraday conversion) for thermal electron distribution function used in this work is provided by \cite{dexter2016public} (see also Equations 33-35 in \citealt{marszewski2021updated}) and reads as follows (notice that $\lambda \equiv 1/\Theta_e$):
\begin{equation}
    \rho_Q = -\frac{n_e e^2 \nu_c^2 \sin^2 (\theta)}{m_e c \nu^3}f_m (X)[\frac{K_1 (\lambda)}{K_2 (\lambda)}+6\lambda^{-1}]
\end{equation}
where
\begin{equation}
\begin{aligned}
    f_m (X) = f_0 (X) + [0.011 \exp (-1.69X^{-1/2}) - 0.003135X^{4/3}] \\ (\frac{1}{2}[1+ \tanh (10\ln (0.6648X^{-1/2}))])
\end{aligned}
\end{equation}
with
\begin{equation}
\begin{aligned}
    f_0 (X) = 2.011 \exp (-19.78X^{-0.5175}) - \\
    \cos (39.89X^{-1/2}) \exp (-70.16X^{-0.6}) - 0.011\exp (-1.69X^{-1/2}).
\end{aligned}
\end{equation}


\bsp	
\label{lastpage}
\end{document}